\documentstyle[aps,prl,twocolumn]{revtex}
\newcommand{\be}{\begin{eqnarray}}
\newcommand{\ee}{\end{eqnarray}}

\begin{document}
\twocolumn[\hsize\textwidth\columnwidth\hsize\csname @twocolumnfalse\endcsname
\draft
\title{Polaron Absorption in a Perovskite Manganite La$_{0.7}$Ca$_{0.3}$MnO$_{3}$}
\author{K. H. Kim, J. H. Jung, and T. W. Noh\cite{Also}}
\address{Department of Physics, Seoul National University, Seoul 151-742, Korea}

\maketitle

\begin{abstract}
~Temperature dependent optical conductivity spectra of
a La$_{{\rm 0.7}}$Ca$_{{\rm 0.3}}$MnO$_{{\rm 3}}$ (LCMO) sample
were measured.
In the metallic regime at very low temperatures, they clearly showed two types of
absorption features, i.e., a sharp Drude peak and a broad mid-infrared absorption
band, which could be
explained as coherent and incoherent bands of a large lattice polaron.
This elementary excitation in LCMO was found to be in a strong coupling
regime and to have interactions with the spin degree of freedom.

\end{abstract}


\vskip2pc]

The recent discovery of colossal magnetoresistance phenomena in doped
perovskite manganites, La${\rm _{1-{\it x}}{\it A}_{{\it x}}MnO_3}$ ({\it A}%
= Sr, Ca) has generated considerable interests. For doping concentrations
with 0.2 $\lesssim ${\it x}$\lesssim $ 0.5, the materials show a transition from a
paramagnetic insulator to a ferromagnetic metal upon cooling near the Curie
temperature {\it T}$_C$. The correlation between metallicity and
ferromagnetism has been explained by the double exchange (DE) model which is
based on the strong Hund coupling between {\it t}$_{2g}$ and {\it e}$_g$
electrons\cite{DE,Kubo}.

In addition to the coupling between the charge and the spin degrees of
freedom, many theoretical\cite{Millis,Zhang} and experimental\cite
{KHKim1,Above,Below,Jaime,Kaplan} works have shown that a coupling between
the charge and the lattice degrees of freedom is quite significant.
Especially, giant oxygen isotope shifts of {\it T}$_C$ in (La,Ca)MnO$_3$
compounds have indicated that the electron-phonon coupling is very large for
the Ca-doped manganites\cite{Zhao}. Therefore, it is widely accepted that a
polaron plays an important role near and above {\it T}$_C$ in the manganites.

Some experimental works demonstrated that local lattice distortions could exist
even in the metallic phase\cite{Below,Louca}. Recently, we observed that the
internal phonon modes of La$_{0.7}$Ca$_{0.3}$MnO$_3$ (LCMO) show significant
frequency shifts, which were explained in terms of changes in electronic
screening during a crossover from a localized polaron regime above {\it T}$%
_C $ to a delocalized one below {\it T}$_C$. For the low temperature ({\it T}%
) state, there are numerous theoretical predictions based on a large polaron
coherent motion\cite{Zhang}, a small polaron tunneling\cite{JDLee}, and a
formation of orbital liquid states\cite{Ishihara}. However, at this moment,
the exact nature of the low {\it T} metallic states is not clearly
understood.

In this letter, we will address polaron absorption features, especially a
very unusual polaron state at low {\it T}, in LCMO. Optical conductivity
spectra at the metallic region showed two types of absorption features, i.e.,
a sharp Drude peak and a broad mid-infrared (IR) absorption band.
Detailed characteristics of the polaron absorption at low {\it T} could be
explained in terms of the large lattice polaron picture by Emin\cite{Emin}.
Moreover, an analysis of the electrodynamics suggested that {\it the large polaron
in LCMO should be in a strong coupling regime}, as far as we know, which has
not been realized in other physical system before.

A polycrystalline LCMO sample was prepared by a standard solid-state
reaction method\cite{anisotropy}. Near normal incidence reflectivity
spectra, $R(\omega )$, were measured between 0.01 and 30 eV. A Fourier
transform spectrophotometer was used for 0.01 $\sim $ 2.5 eV and a grating
type monochromator was used between 0.4 and 7.0 eV. Above 6.0 eV, the
synchrotron radiation source from the Normal Incidence Monochromator beam
line at Pohang Light Source (PLS) was used\cite{JHJung}. ~To measure {\it %
T}-dependent $R(\omega )$ below 2.5 eV, a liquid He-cooled cryostat was used.

Optical conductivity spectra, $\sigma $($\omega $), were obtained using the
Kramers-Kronig transformation. For this analysis, $R(\omega )$ below 0.01 eV
were extrapolated with the Hagen-Rubens relation\cite{KHKim2}. For a high
frequency region, reflectivity, $R$, at 30 eV was extended up to 40 eV,
above which $\omega ^{-4}$ dependence was assumed. We found that there were
less than 2.0 \% changes of $R$ with {\it T} in a frequency region between
2.0 and 2.5 eV, so we attached low temperature $R$ data below 2.5 eV
smoothly with room temperature data above it. The errors due to such an
extrapolation were estimated to be about 10 \% in $\sigma $($\omega $)
around 2.5 eV and smaller below 2.0 eV.

Fig. 1 (a) shows that there are significant spectral weight (SW) transfers
from high to low energies with decreasing {\it T}. A crossover energy is
about 0.5 eV. It is noted that the spectra below 0.5 eV, shown in Fig. 1
(b), can be characterized by two types of responses, i.e., a sharp Drude
peak in a far-IR region and a broad absorption band in a mid-IR region. So,
a corresponding conductivity spectrum can be written as a sum of the two
contributions: $\sigma $($\omega $)=$\sigma _{Drude}$($\omega $)+$\sigma
_{MIR}$($\omega $). A more clear picture on the development of the far-IR\
Drude peak is also shown in the inset of Fig. 1 (b). As {\it T} decreases, $%
\sigma _{MIR}$($\omega $) increases initially but becomes saturated around
120 K. However, $\sigma _{Drude}$($\omega $) increases continuously without
any saturation. The behaviors of $\sigma $($\omega $) in LCMO are quite
different from those in a Nd$_{0.7}$Sr$_{0.3}$MnO$_3$ film\cite{Kaplan},
which showed no Drude peak and a strong and quite symmetric mid-IR band. On
the other hand, the general behaviors of $\sigma $($\omega $) in LCMO are
somewhat similar to those in a La$_{0.7}$Sr$_{0.3}$MnO$_3$ single crystal\cite
{Okimoto}. However, for La$_{0.7}$Sr$_{0.3}$MnO$_3$, quantitative
information of both $\sigma _{Drude}$($\omega $) and $\sigma _{MIR}$($%
\omega $) was still lacking, and the origin of the mid-IR band has not been
clearly explained yet\cite{Okimoto}.

Above {\it T}$_C$, i.e., in the insulating regime, it is widely accepted
that a small polaron, called the ``Holstein polaron'', plays an important
role. And, $\sigma _{MIR}$($\omega $) below 0.8 eV at 260 K can be
fitted reasonably well with the small polaron model\cite{Reik,Mott}, just
like the case of (La,Sr)$_2$NiO$_{4+\delta}$\cite{Eklund}. However, below {\it T}$_C$%
, it is not known clearly whether a large polaron coherent motion\cite{Zhang}
or small polaron tunneling\cite{JDLee} will be the origin of the metallic
behavior in LCMO.

Emin investigated frequency responses of the large and the small polaron
absorption\cite{Emin}. A coherent band of the large polaron should show up
only at lower frequencies below characteristic phonon modes and become more
significant as {\it T} decreases. And, a photoionization of the large
polaron should bring out an incoherent mid-IR band which is very asymmetric
and shows a long tail above its peak position. On the other hand, the
coherent band of small polaron, i.e., polaron tunneling band, should occur at
an energy region much lower than the characteristic phonon modes.
Behaviors of both $\sigma _{Drude}$($\omega $) and $\sigma _{MIR}$($\omega $%
) at {\it T} $\lesssim $120 K, shown in Fig. 1, are quite consistent with
the large polaron absorption features, predicted by Emin. As shown in the
inset, the Drude peak becomes evident below the bending phonon mode
frequency, i.e., $\sim$ 330 cm$^{-1}$, and its width does not change too
much. Moreover, $\sigma _{MIR}$($\omega $) near 20 K is quite asymmetric
and shows a long tail above its peak position. From this observation, $%
\sigma _{Drude}$($\omega $) at {\it T} $\lesssim $120 K can be attributed to
a coherent motion of a large polaron, and $\sigma _{MIR}$($\omega $) can be
attributed to its incoherent absorption band\cite{Local}.

To get quantitative information on the electrodynamic responses of the large
polaron in LCMO, we looked into the real part of the dielectric function, $%
\varepsilon _1$. Figure 2 shows {\it T-}dependent $\varepsilon _1$ spectra.
At a low frequency $\varepsilon _1$ becomes negative, indicating that LCMO
becomes metallic. In a simple Drude model, the complex dielectric function
$\widetilde{\varepsilon }(\omega )$ can be written as
\begin{equation}
\widetilde{\varepsilon }(\omega )=\varepsilon _\infty -\ \frac{\omega _p^2}{%
\omega ^2+i\omega /\tau },
\end{equation}
where $\omega _p$ and $\tau $ are the bare plasma frequency and the
relaxation time of free carriers, respectively. And, $\varepsilon
_\infty $ is the dielectric constant at a high frequency. If $\omega \gg
1/\tau $, $\widetilde{\varepsilon }(\omega )\approx $ $\varepsilon _\infty
-\ \omega _p^2/\omega ^2$. In a real metallic sample, however, mid-IR and
interband absorptions also contribute to $\widetilde{\varepsilon }(\omega )$%
. If their contributions to $\widetilde{\varepsilon }(\omega )$ vary slowly
in $\omega $, $\varepsilon _1$ can be approximated as $\varepsilon _h-\
\omega _p^2/\omega ^2$, where $\varepsilon _{h\text{ }}$represents a
``background'' dielectric constant at a high frequency determined by $%
\varepsilon _\infty $ and contributions from the interband and the mid-IR
absorptions. Then, the slope in a $\varepsilon _{1\text{ }}$vs $\omega
^{-2}$ plot will provide the value of $\omega _p^2$ for the coherent band.
The inset of Fig. 2 shows the $\varepsilon _{1\text{ }}$vs $\omega ^{-2}$
plot. [$\varepsilon _h$ was estimated to be about 4.9.] The solid and dotted
lines are experimental data and linear guide lines, respectively. Except
phonon frequency regions, the $\varepsilon _{1\text{ }}$vs $\omega ^{-2}$
plots are quite linear.

The experimental values of $\omega _p^2$ are plotted in Fig. 3 (a).
Interestingly, $\omega _p^2$ is approximately proportional to ({\it T}$_{%
{\rm {C}}}$-{\it T}). The increase of $\omega _p^2$ upon cooling is in good
agreement with the photoemission data that density of states at Fermi energy increases
progressively below {\it T}$_C$\cite{JHPark}. Note that $\omega _p^2=4\pi
ne^2/m^{*}$, where {\it n} and {\it m*} represent a density and an effective
mass of the free carriers, respectively. In the large lattice polaron
picture, $m^{*}=m_b+M_p$, where $m_b$ is the band mass of the carriers and $%
M_p$ represents the atomic contribution due to the shift of the equilibrium
position of each atom when the polaron moves by a lattice constant\cite
{Emin,Mahan}.

To get a further insight on the coherent polaron motion in LCMO, we need
information on its dc resistivity, $\rho $. Since the $\rho $ data for a
LCMO single crystal are not available, they were estimated from the{\it \ R}
data using the Hagen-Rubens relation\cite{KHKim2,SumRule}. As shown in Fig.
3 (b), $\rho $ increases rapidly above 150 K. Using a relation such that $%
\rho =m^{*}/ne^2\tau $, the scattering rate $1/\tau $ for the Drude peak was
estimated and plotted in Fig. 3 (c). Another interesting physical quantity
related to the coherent polaron motion is its mean free path, {\it l }[=$%
\tau \hbar $ (3$\pi ^2n$)$^{1/3}$/$m^{*}$]. If we assume that {\it n} is
equal to 0.3 hole per Mn at 20 K, the $\omega _p^2$ value predicts {\it m}$%
^{*}$ as 13{\it m}$_e$. Using the experimental values of $\tau $, {\it l }%
could be estimated. At 20 K, {\it l} $\sim $ 25 \AA\ which is much larger
than the lattice constant of about 3.9 \AA . However, as {\it T} increases,
{\it l} becomes smaller and approaches the lattice constant value near {\it T%
}$_C$: {\it l} $\sim $ 4.2 \AA\ at 240 K. Note that this result is
consistent with the Ioffe-Regel criterion for metal-insulator transitions,
i.e. {\it l} $\sim $(lattice constant)\cite{Rao}.

The solid rectangles and the solid circles in Fig. 4 represent the spectral
weights of the Drude peak and the mid-IR band, respectively. [The open
diamonds represent the total SW.] The Drude weight (DW) of the coherent
polaron motion was evaluated using $(2m_e/\pi e^2N)$($\omega _p^2/8)$, where
$m_e$ is an electron mass and {\it N} is the number of Mn atoms per unit
volume. Then, total effective carriers below a cut-off energy, $%
N_{eff}(\omega _c)$, can be estimated from
\begin{equation}
N_{eff}(\omega _c)=\frac{2m_e}{\pi e^2N}\int_0^{\omega _c}\sigma (\omega
)d\omega \text{ .}
\end{equation}
We chose the cut-off energy $\hbar \omega _{c\text{ }}$as 0.5 eV. By
subtracting the DW from $N_{eff}(\omega _c)$, the spectral weight of the
mid-IR band (SWMB) was obtained. As {\it T} decreases, the SWMB increases
but saturates at low temperatures. On the other hand, the DW can be scaled
with ({\it T}$_{{\rm {C}}}$-{\it T}). The DW comprises a small portion of $%
N_{eff}(\omega _c)$ below {\it T}$_C$: at 240K, DW$\approx $ $%
0.07N_{eff}(\omega _c)$ and even at 20 K, DW$\approx $ $0.33N_{eff}(\omega
_c)$. The values of the DW are very small at overall temperatures, compared
with the doped carrier density, i. e. 0.3 hole per Mn. The small values of
the DW might come from the large effective mass of the coherent polaron
motion especially at low {\it T}. In case of a strongly coupled large
polaron, $M_p\approx 0.02m_b\alpha ^4$, where $\alpha $ represents the
Fr\"{o}hlich coupling constant\cite{Mahan}. If we assume $m_b$=$m_e$, $%
\alpha $ is estimated to about 5, which suggests that the large polaron in
LCMO should remain in a strong coupling regime.

In literature, the realization of the strongly coupled large polaron was
questioned due to its screening effect arising from dense carrier
concentrations\cite{Mott}. In La$_{0.7}$Ca$_{0.3}$MnO$_3$, there are lots of
Mn$^{3+}$ sites which are susceptible to the local Jahn-Teller distortion,
so screening effect could be very important. However, in the DE model, a
motion of a carrier at the metallic regime can be affected by the spin
ordering\cite{SpinPolaron}. Therefore, the lattice at low {\it T} can be
also influenced by the long range spin ordering\cite{Above}. Under this
influence, the coupling between the electron and the lattice in LCMO could
be extended beyond a single lattice site. A recent pulsed neutron
diffraction study also shows that the lattice polaron becomes more extended
at low {\it T}\cite{Louca}.

Our experimental data also clearly indicate that {\it the lattice polaron
state in LCMO is also coupled with the spin degree of freedom}. According to
the large lattice polaron theory by Emin\cite{Emin}, 1/$\tau $ should be
linearly proportional to {\it T}. Fig. 3 (c) shows that 1/$\tau $ for the
coherent polaron band remains nearly constant below 120 K and starts to
increase above 160 K. In the DE picture, as {\it T }approaches to {\it T}$_C$%
, the free carriers are more likely to be scattered by the spin fluctuation.
In the mean field limit, such a spin alignment effect could be represented
by $\gamma _{\text{DE}}({\it T})$ $\equiv $
\mbox{$<$}
cos($\theta $/2)%
\mbox{$>$}
, which is a thermodynamic average of cos($\theta$/2), where $\theta $ is
an angle between nearest neighbor spins\cite{Kubo}. Shown as a solid line in
Fig. 3 (c), 1/$\tau $ can be approximately scaled with 1/$\gamma _{\text{DE}}(%
{\it T})$. Moreover, it is very interesting to see that the {\it T}%
-dependence of the SWMB is quite close to $\gamma _{\text{DE}}$({\it T}),
shown as the solid line in Fig. 4. Therefore, it is highly likely that both
the coherent and the incoherent polaron bands are related to the spin
degree of freedom. Note that a similar coupling existed in the phonon
frequency renormalization\cite{KHKim1} through $\widetilde{\omega }_q\simeq
\omega _q(1-\beta /\gamma _{\text{DE}}$($T$)$)^{1/2}$.

In summary, we investigated the optical responses of a perovskite manganite La$%
_{0.7} $Ca$_{0.3}$MnO$_3$. They showed the Drude peak and the broad
mid-infrared absorption band. At low temperature, these absorption features
were interpreted as coherent and incoherent bands of a large lattice
polaron. The electrodynamic analysis showed that the polaron was in a strong
coupling regime, which is quite unique in this manganite. Moreover, the
coherent and incoherent bands seem to have strong correlations with the spin
degree of freedom.

We thank for useful discussions with Prof. E. J. Choi, Prof. J. Yu, Prof. B.
I. Min, and Prof. H. Y. Choi. This work was financially supported by the
Ministry of Education through the Basic Science Research Institute program
BSRI-97-2416, and by KOSEF through grant No. 96-0702-02-01-3. Experiments at
PLS were supported by MOST and POSCO.

\begin{figure}[tbp]
\caption{(a) Optical conductivity spectra below 2 eV. (b) Optical
conductivity spectra below 0.5 eV. Temperature ranges are the same as those
in (a). Inset shows detailed optical conductivity spectra in a far-IR
region. }
\end{figure}

\begin{figure}[tbp]
\caption{The real part of the complex dielectric function, $\varepsilon _1,$
spectra. Inset shows the $\varepsilon _1$vs $\omega ^{-2}$ plots at various
temperatures. Temperature ranges are the same with those shown in the inset
except 260 K. }
\label{real dielectric}
\end{figure}
\begin{figure}[tbp]
\caption{(a) Values of $\omega _p^2$ determined from the $\varepsilon _1$vs $%
\omega ^{-2}$ plots. The dotted line is a linear guide one. (b) Resistivity
values determined from the Hagen-Rubens relation. (c) Scattering rates of
the free carriers determined from (a) and (b). The behavior of 1/$\gamma _{%
\text{DE}}$({\it T}) is overlapped as a solid line.}
\end{figure}
\begin{figure}[tbp]
\caption{$T$-dependence of N$_{eff}$ (0.5 eV) (open diamonds), Drude weight
(solid rectangles), and spectral weight of the mid-IR band (solid circles)
in La$_{{\rm 0.7}}$Ca$_{{\rm 0.3}}$MnO$_{{\rm 3}}$. A solid line represents
the behavior of $\gamma _{\text{DE}}$({\it T}). A dotted line is a linear
guide one.}
\label{Model}
\end{figure}


\begin{references}
\bibitem{SpinPolaron}  Neglecting the lattice distortion effects, such a coupling
between electron and spin degrees of freedom can result in an elementary
excitation, called a ``spin polaron''.
\end{references}
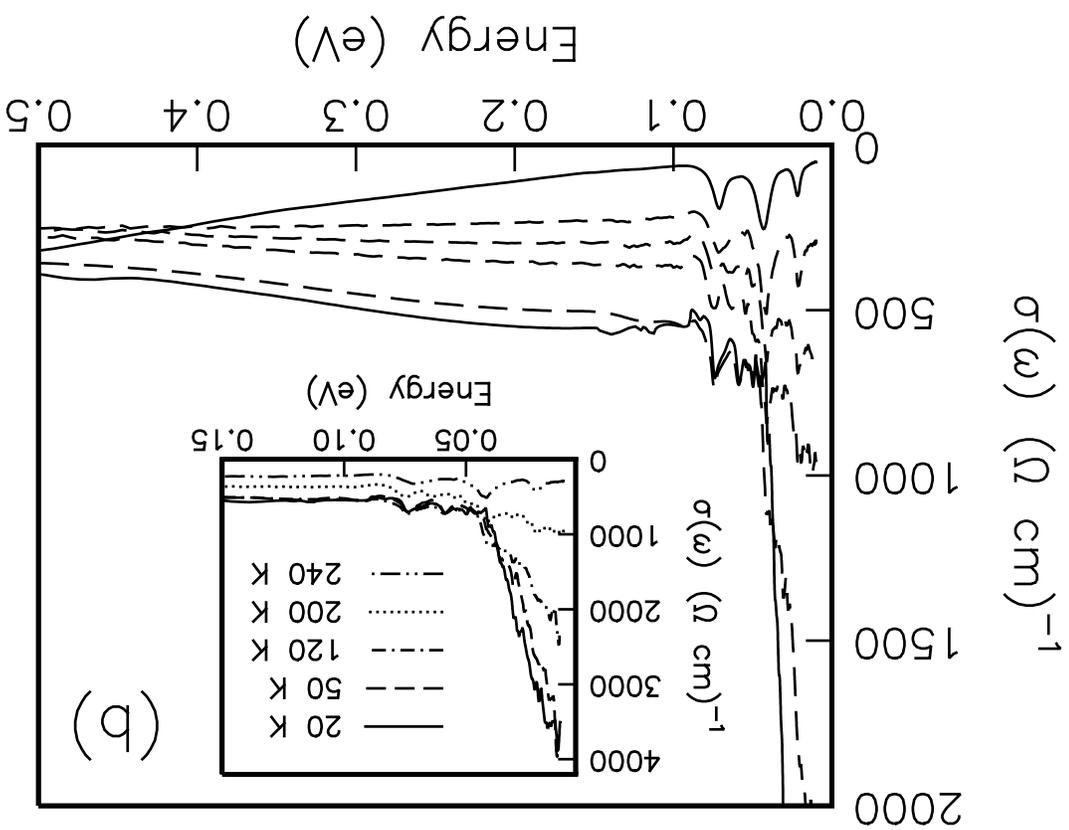
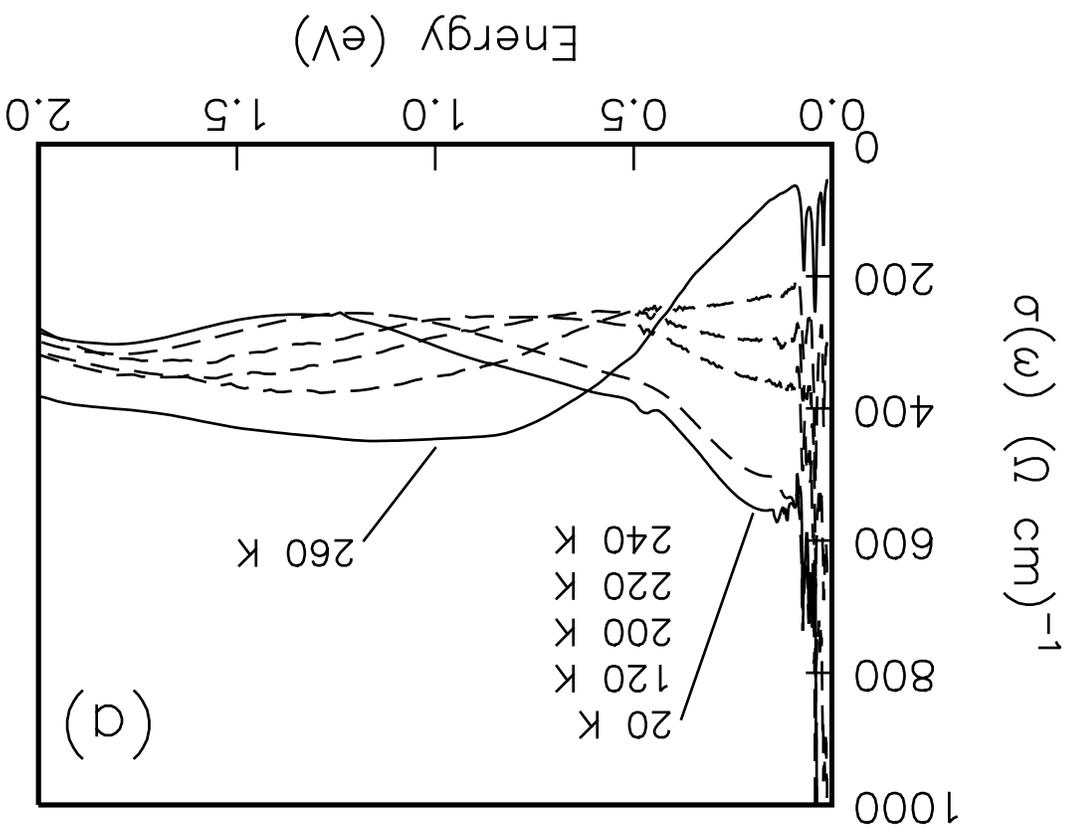
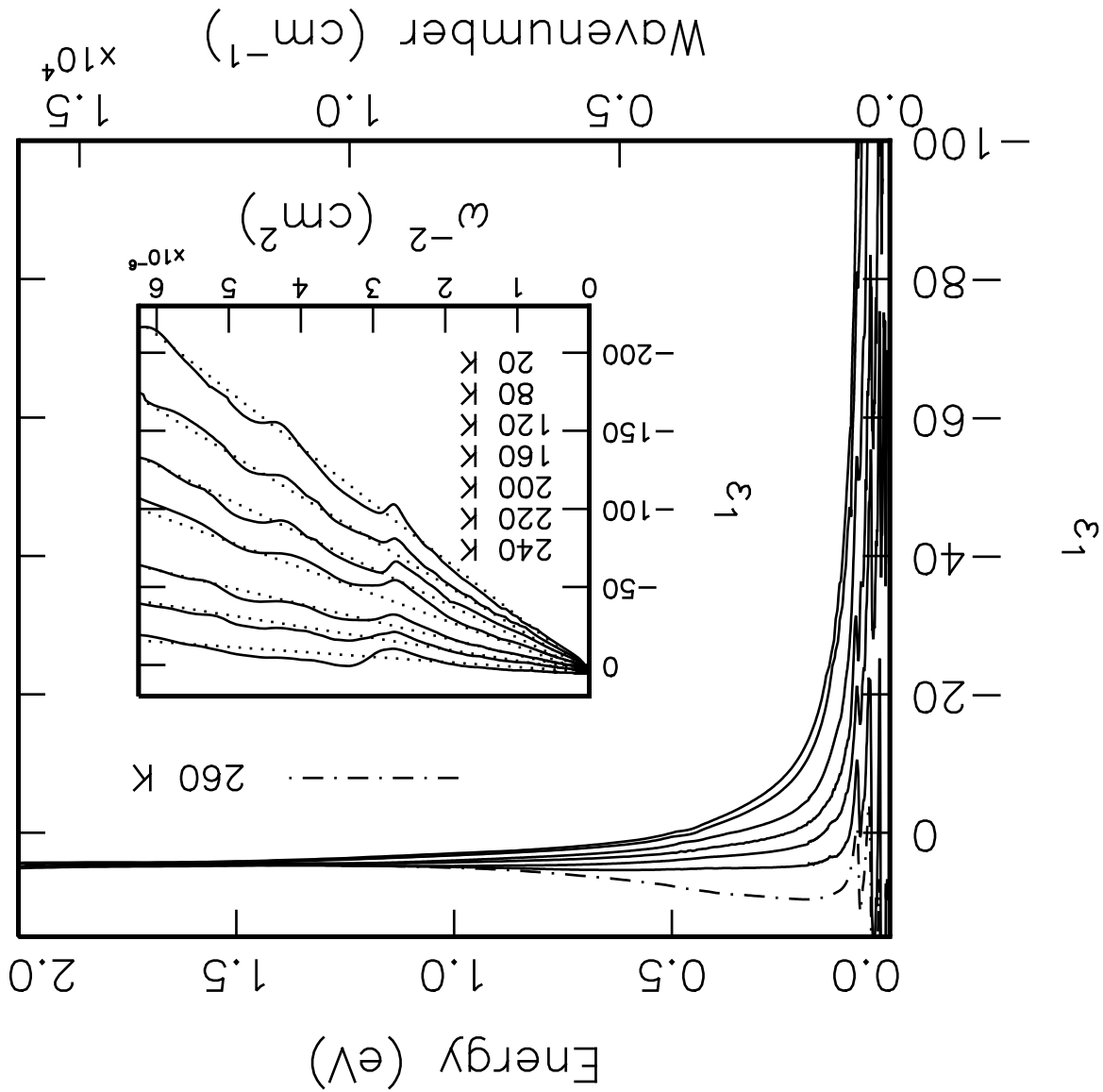
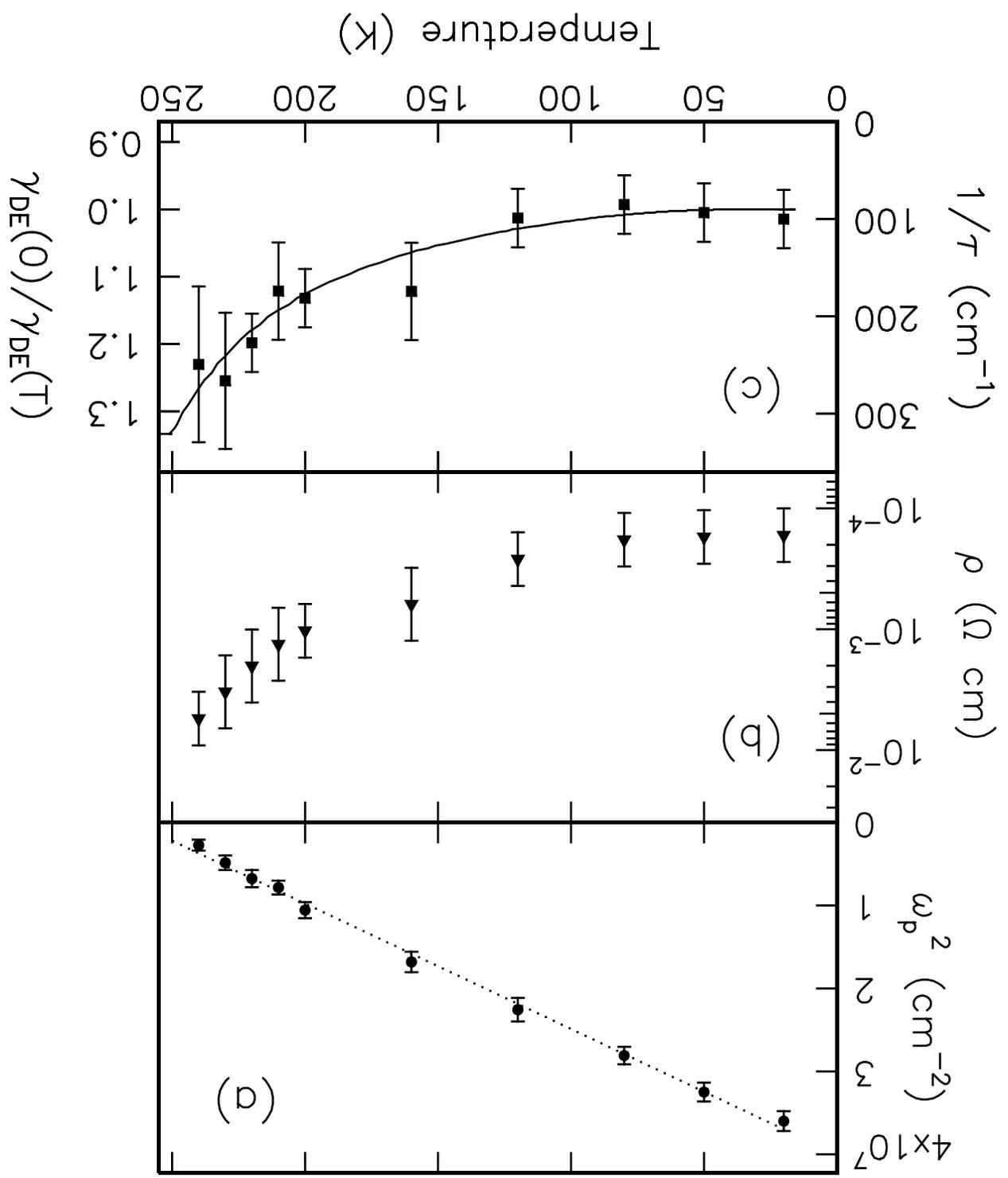
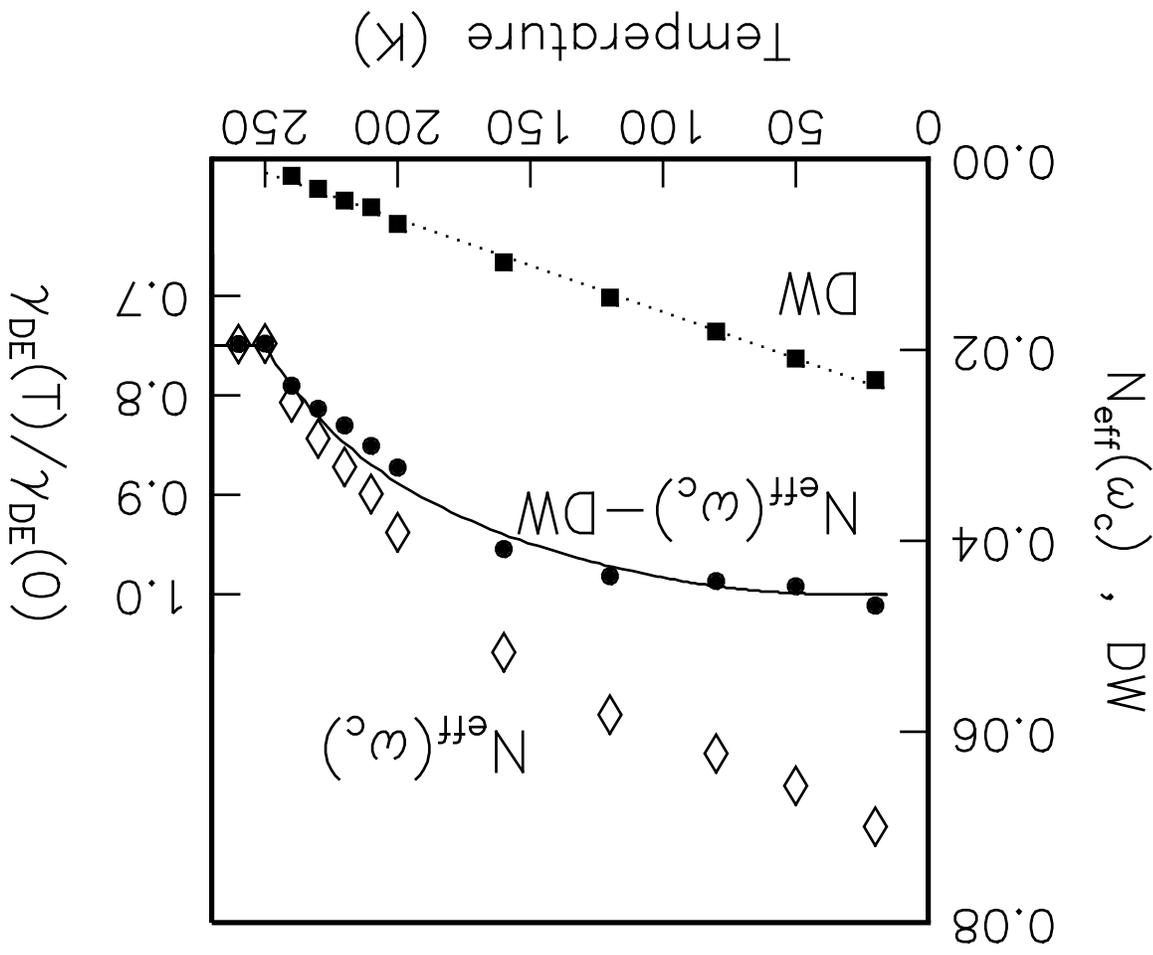
\end{document}